\documentclass[a4paper]{article}
\usepackage{INTERSPEECH2018}
\usepackage[utf8]{inputenc}
\usepackage{multirow}
\usepackage{balance}
\usepackage[normalem]{ulem}

\newcommand{\citep}[1]{\cite{#1}}
\newcommand{\citet}[1]{\cite{#1}}

\usepackage{color}

\newcommand{\ve}[1]{\mathbf{#1}}                

\title{Self-Attention Linguistic-Acoustic Decoder}
\name{Santiago Pascual$^1$, Antonio Bonafonte$^1$, Joan Serr\`a$^2$}
\address{
  $^1$Universitat Polit\`ecnica de Catalunya, Barcelona, Spain\\
  $^2$Telef\'onica Research, Barcelona, Spain}
\email{santi.pascual@upc.edu}

\begin{document}

\maketitle
\begin{abstract}
The conversion from text to speech relies on the accurate mapping from linguistic to acoustic symbol sequences, for which current practice employs recurrent statistical models like recurrent neural networks. Despite the good performance of such models (in terms of low distortion in the generated speech), their recursive structure tends to make them slow to train and to sample from. In this work, we try to overcome the limitations of recursive structure by using a module based on the transformer decoder network, designed without recurrent connections but emulating them with attention and positioning codes. Our results show that the proposed decoder network is competitive in terms of distortion when compared to a recurrent baseline, whilst being significantly faster in terms of CPU inference time. On average, it increases Mel cepstral distortion between 0.1 and 0.3\,dB, but it is over an order of magnitude faster on average. Fast inference is important for the deployment of speech synthesis systems on devices with restricted resources, like mobile phones or embedded systems, where speaking virtual assistants are gaining importance.
\end{abstract}

\section{Introduction}

Speech synthesis makes machines generate speech signals, and text-to-speech (TTS) conditions speech generation on input linguistic contents. Current TTS systems use statistical models like deep neural networks to map linguistic/prosodic features extracted from text to an acoustic representation. This acoustic representation typically comes from a vocoding process of the speech waveforms, and it is decoded into waveforms again at the output of the statistical model~\cite{zen2015acoustic}.

To build the linguistic to acoustic mapping, deep network TTS models make use of a two-stage structure~\cite{ze2013statistical}. The first stage predicts the number of frames (duration) of a phoneme to be synthesized with a duration model, whose inputs are linguistic and prosodic features extracted from text. In the second stage, the acoustic parameters of every frame are estimated by the so-called acoustic model. Here, linguistic input features are added to the phoneme duration predicted in the first stage. Different works use this design, outperforming previously existing statistical parametric speech synthesis systems with different variants in prosodic and linguistic features, as well as perceptual losses of different kinds in the acoustic mapping~\cite{lu2013combining, qian2014training, hu2015fusion, hu2014investigation, kang2013multi}. Since speech synthesis is a sequence generation problem, recurrent neural networks (RNNs) are a natural fit to this task. They have thus been used as deep architectures that effectively predict either prosodic features~\cite{pascual2016prosodic, chen1998rnn} or duration and acoustic features~\cite{achanta2015investigation,wuinvestigating,fernandez2014prosody,zen2015unidirectional,pascual2016multi}. Some of these works also investigate possible performance differences using different RNN cell types, like long short-term memory (LSTM)~\cite{hochreiter1997long} or gated recurrent unit~\cite{chung2014empirical} modules. 

In this work, we propose a new acoustic model, based on part of the Transformer network~\cite{vaswani2017attention}. 
The original Transformer was designed as a sequence-to-sequence model for machine translation. Typically, in sequence-to-sequence problems, RNNs of some sort were applied to deal with the conversion between the two sequences~\cite{sutskever2014sequence,bahdanau2014neural}. The Transformer substitutes these recurrent components by attention models and positioning codes that act like time stamps. In~\cite{vaswani2017attention}, they specifically introduce the self-attention mechanism, which can relate elements within a single sequence without an ordered processing (like RNNs) by using a compatibility function, and then the order is imposed by the positioning code. The main part we import from that work is the encoder, as we are dealing with a mapping between two sequences that have the same time resolution. We however call this part the decoder in our case, given that we are decoding linguistic contents into their acoustic codes. We empirically find that this Transformer network is as competitive as a recurrent architecture, but with faster inference/training times.


This paper is structured as follows. In section~\ref{sec:satt}, we describe the self-attention linguistic-acoustic decoder (SALAD) we propose. Then, in section~\ref{sec:exp_setup}, we describe the followed experimental setup, specifying the data, the features, and the hyper-parameters chosen for the overall architecture. Finally, results and conclusions are shown and discussed in sections~\ref{sec:results} and~\ref{sec:conclusions}, respectively. The code for the proposed model and the baselines can be found in our public repository~\footnote{https://github.com/santi-pdp/musa\_tts}.




\section{Self-Attention Linguistic-Acoustic Decoder}
\label{sec:satt}

To study the introduction of a Transformer network into a TTS system, we employ our previous multiple speaker adaptation (MUSA) framework~\cite{pascual2016multi, pascual2016deep, pascual2016alpha}. This is a two-stage RNN model influenced by the work of Zen~and~Sak~\cite{zen2015unidirectional}, in the sense that it uses unidirectional LSTMs to build the duration model and the acoustic model without the need of predicting dynamic acoustic features. A key difference between our works and~\cite{zen2015unidirectional} is the capacity to model many speakers and adapt the acoustic mapping among them with different output branches, as well as interpolating new voices out of their common representation. Nonetheless, for the current work, we did not use this multiple speaker capability and focused on just one speaker for the new architecture design on improving the acoustic model. 

\begin{figure*}[t]
\centering
\includegraphics[width=0.7\linewidth]{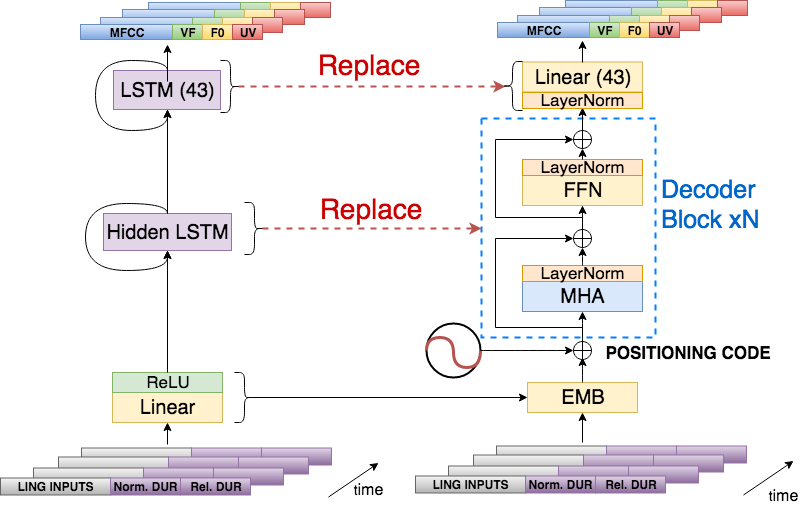}
\caption{\label{fig:rnn_vs_salad} Transition from RNN/LSTM acoustic model to SALAD. The embedding projections are the same. Positioning encoding introduces sequential information. The decoder block is stacked $N$ times to form the whole structure replacing the recurrent core. FFN: Feed-forward Network. MHA: Multi-Head Attention.}
\end{figure*}

The design differences between the RNN and the transformer approaches are depicted in figure~\ref{fig:rnn_vs_salad}. In the MUSA framework with RNNs, we have a pre-projection fully-connected layer with a ReLU activation that reduces the sparsity of linguistic and prosodic features. This embeds the mixture $\ve{x}_t$ of different input types into a common representation $\ve{h}_t$ in the form of one vector per time step $t$. Hence, the transformation $\mathbb{R}^{L} \rightarrow \mathbb{R}^{H}$ is applied independently at each  time step $t$ as 
\begin{equation*}
\label{eq:emb}
\ve{h}_{t} = \max(0, \ve{W}\ve{x}_{t} + \ve{b}),
\end{equation*}
where $\ve{W} \in \mathbb{R}^{H\times L}$, $\ve{b} \in \mathbb{R}^H$, $\ve{x}_t \in \mathbb{R}^L$, and $\ve{h}_t \in \mathbb{R}^H$. After this projection, we have the recurrent core formed by an LSTM layer of size $H$ and an additional LSTM output layer. The MUSA-RNN output is recurrent, as this prompted better results than using dynamic features to smooth cepstral trajectories in time~\cite{zen2015unidirectional}.

Based on the Transformer architecture~\cite{vaswani2017attention}, we propose a pseudo-sequential processing network that can leverage distant element interactions within the input linguistic sequence to predict acoustic features. This is similar to what an RNN does, but discarding any recurrent connection. This will allow us to process all input elements in parallel at inference, hence substantially accelerating the acoustic predictions. In our setup, we do not face a sequence-to-sequence problem as stated previously, so we only use a structure like the Transformer encoder which we call a linguistic-acoustic decoder. 

The proposed SALAD architecture begins with the same embedding of linguistic and prosodic features, followed by a positioning encoding system. As we have no recurrent structure, and hence no processing order, this positioning encoding system will allow the upper parts of the network to locate their operating point in time, such that the network will know where it is inside the input sequence~\cite{vaswani2017attention}. This positioning code $\ve{c}\in\mathbb{R}^H$ is a combination of harmonic signals of varying frequency:
\begin{equation*}
\begin{aligned}
\label{eq:pos_code}
&c_{t, 2i} = \sin\left(t/10000^\frac{2i}{H}\right)\\
&c_{t, 2i+1} = \cos\left(t/10000^\frac{2i}{H}\right)
\end{aligned}
\end{equation*}
where $i$ represents each dimension within $H$. At each time-step $t$, we have a unique combination of signals that serves as a time stamp, and we can expect this to generalize better to long sequences than having an incremental counter that marks the position relative to the beginning. Each time stamp $\ve{c_t}$ is summed to each embedding $\ve{h_t}$, and this is input to the decoder core.

The decoder core is built with a stack of $N$ blocks, depicted within the dashed blue rectangle in figure~\ref{fig:rnn_vs_salad}. These blocks are the same as the ones proposed in the decoder of~\cite{vaswani2017attention}, but we only have self-attention modules to the input, so it looks more like the Transformer encoder. The most salient part of this type of block is the multi-head attention (MHA) layer. This applies $h$ parallel self-attention layers, which can have a more versatile feature extraction than a single attention layer with the possibility of smoothing intra-sequential interactions. After the MHA comes the feed-forward network (FFN), composed of two fully-connected layers. The first layer expands the attended features into a higher dimension $d_{\text{ff}}$, and this gets projected again to the embedding dimensionality $H$. Finally, the output layer is a fully-connected dimension adapter such that it can convert the hidden dimensions $H$ to the desired amount of acoustic outputs, which in our case is 43 as discussed in section~\ref{sec:ling_aco_feats}. As stated earlier, we may slightly degrade the quality of predictions with this output topology, as recurrence helps in the output layer capturing better the dynamics of acoustic features. Nonetheless, this can suffice our objective of having a highly parallelizable and competitive system.

\section{Experimental Setup}
\label{sec:exp_setup}

\subsection{Dataset}

For the experiments we use utterances of speakers from the TCSTAR project dataset~\cite{bonafonte2006tc}. This corpora includes sentences and paragraphs taken from transcribed parliamentary speech and transcribed broadcast news. The purpose of these text sources is twofold: enrich the vocabulary and facilitate the selection of the sentences to achieve good prosodic and phonetic coverage. For this work, we choose the same male (M1) and female (F1) speakers as in our previous works. These two speakers have the most amount of data among the available ones. Their amount of data is balanced with approximately the following durations per split for both: 100 minutes for training, 15 minutes for validation, and 15 minutes for test.

\subsection{Linguistic and Acoustic Features}
\label{sec:ling_aco_feats}
The decoder maps linguistic and prosodic features into acoustic ones. This means that we first extract hand-crafted features out of the input textual query. These are extracted in the label format, following our previous work in~\cite{pascual2016deep}. We thus have a combination of sparse identifiers in the form of one-hot vectors, binary values, and real values. These include the identity of phonemes within a window of context, part of speech tags, distance from syllables to end of sentence, etc. For more detail we refer to~\cite{pascual2016deep} and references therein. 

For a textual query of $N$ words, we will obtain $M$ label vectors, $M \geq N$, each with 362 dimensions. In order to inject these into the acoustic decoder, we need an extra step though. As mentioned, the MUSA testbed follows the two-stage structure: (1) duration prediction and (2) acoustic prediction with the amount of frames specified in first stage. Here we are only working with the acoustic mapping, so we enforce the duration with labeled data. For this reason, and similarly to what we did in previous works~\cite{pascual2016multi,pascual2016alpha}, we replicate the linguistic label vector of each phoneme as many times as dictated by the ground-truth annotated duration, appending two extra dimensions to the 362 existing ones. These two extra dimensions correspond to (1) absolute duration normalized between 0 and 1, given the training data, and (2) relative position of current phoneme inside the absolute duration, also normalized between 0 and 1. 

We parameterize the speech with a vocoded representation using Ahocoder~\cite{erro2011improved}. Ahocoder is an harmonic-plus-noise high quality vocoder, which converts each windowed waveform frame into three types of features: (1) mel-frequency cepstral coefficients (MFCCs), (2) log-F0 contour, and (3) voicing frequency (VF). Note that F0 contours have two states: either they follow a continuous envelope for voiced sections of speech, or they are 0, for which the logarithm is undefined. Because of that, Ahocoder encodes this value with $-10^{9}$, to avoid numerical undefined values. This result would be a cumbersome output distribution to be predicted by a neural net using a quadratic regression loss. Therefore, to smooth the values out and normalize the log-F0 distribution, we linearly interpolate these contours and create an extra acoustic feature, the unvoiced-voiced flag (UV), which is the binary flag indicating the voiced or unvoiced state of the current frame. We will then have an acoustic vector with 40~MFCCs, 1~log-F0, 1~VF, and 1~UV. This equals a total number of 43~features per frame, where each frame window has a stride of 80 samples over the waveform. Real-numbered linguistic features are Z-normalized by computing statistics on the training data. In the acoustic feature outputs, all of them are normalized to fall within $[0, 1]$.

\subsection{Model Details and Training Setup}
\label{sec:model_details}
We have two main structures: the baseline MUSA-RNN and SALAD. The RNN takes the form of an LSTM network for their known advantages of avoiding typical vanilla RNN pitfalls in terms of vanishing memory and bad gradient flows. Each of the two different models has two configurations, small (Small RNN/Small SALAD) and big (Big RNN/Big SALAD). This intends to show the performance difference with regard to speed and distortion between the proposed model and the baseline, but also their variability with respect to their capacity (RNN and SALAD models of the same capacity have an equivalent number of parameters although they have different connexion topologies). Figure~\ref{fig:rnn_vs_salad} depicts both models' structure, where only the size of their layers (LSTM, embedding, MHA, and FFN) changes with the mentioned magnitude. Table~\ref{tab:models_magnitudes} summarizes the different layer sizes for both types of models and magnitudes.

\begin{table}[t]
  \caption{Different layer sizes of the different models. Emb: linear embedding layer, and hidden size $H$ for SALAD models in all layers but FFN ones. HidRNN: Hidden LSTM layer size. $d_{\text{ff}}$: Dimension of the feed-forward hidden layer inside the FFN.}
  \label{tab:models_magnitudes}
  \centering
  \setlength{\tabcolsep}{5pt}
  \begin{tabular}{lcccc}
    \toprule
    \textbf{Model} & \textbf{Emb} & \textbf{HidRNN} & $\textbf{d}_{\text{ff}}$ \\
    \midrule
    Small RNN                       & 128 & 450 &  -             \\
    Small SALAD                     & 128 & - & 1024 		\\
    Big RNN                       & 512 & 1300  & -					\\
    Big SALAD					   & 512 & - & 2048       \\
    \bottomrule
  \end{tabular}
\end{table}

Both models have dropout~\cite{srivastava2014dropout} in certain parts of their structure. The RNN models have it after the hidden LSTM layer, whereas the SALAD model has many dropouts in different parts of its submodules, replicating the ones proposed in the original Transformer encoder~\cite{vaswani2017attention}. The RNN dropout is 0.5, and SALAD has a dropout of 0.1 in its attention components and 0.5 in FFN and after the positioning codes.

Concerning the training setup, all models are trained with batches of 32 sequences of 120 symbols. The training is in a so-called stateful arrangement, such that we carry the sequential state between batches over time (that is, the memory state in the RNN and the position code index in SALAD). To achieve this, we concatenate all the sequences into a very long one and chop it into 32 long pieces. We then use a non-overlapped sliding window of size 120, so that each batch contains a piece per sequence, continuous with the previous batch. This makes the models learn how to deal with sequences longer than 120 outside of train, learning to use a conditioning state different than zero in training. Both models are trained for a maximum of 300 epochs, but they trigger a break by early-stopping with the validation data. The validation criteria for which they stop is the mel cepstral distortion (MCD; discussed in section~\ref{sec:results}) with a patience of 20 epochs.

Regarding the optimizers, we use Adam~\cite{kingma2014adam} for the RNN models, with the default parameters in PyTorch ($\text{lr}=0.001$, $\beta_1=0.9$, $\beta_2=0.999$, and $\epsilon=10^{-8}$). For SALAD we use a variant of Adam with adaptive learning rate, already proposed in the Transformer work, called Noam~\cite{vaswani2017attention}. This optimizer is based on Adam with $\beta_1=0.9$, $\beta_2=0.98$, $\epsilon=10^{-9}$ and a learning rate scheduled with
\begin{equation*}
\text{lr} = H^{-0.5} \cdot \min(s^{-0.5}, s \cdot w^{-1.5})
\end{equation*}
where we have an increasing learning rate for $w$ warmup training batches, and it decreases afterwards, proportionally to the inverse square root of the step number $s$ (number of batches). We use $w=4000$ in all experiments. The parameter $H$ is the inner embedding size of SALAD, which is $128$ or $512$ depending on whether it is the small or big model as noted in table~\ref{tab:models_magnitudes}. We also tested Adam on the big version of SALAD, but we did not observe any improvement in the results, so we stick to Noam following the original Transformer setup.

\section{Results}
\label{sec:results}

In order to assess the distortion introduced by both models, we took three different objective evaluation metrics. First, we have the MCD measured in decibels, which tells us the amount of distortion in the prediction of the spectral envelope. Then we have the root mean squared error (RMSE) of the F0 prediction in Hertz. And finally, as we introduced the binary flag that specifies which frames are voiced or unvoiced, we measure the accuracy (number of correct hits over total outcomes) of this binary classification prediction, where classes are balanced by nature. These metrics follow the same formulations as in our previous works~\cite{pascual2016multi,pascual2016deep,pascual2016alpha}.

\begin{figure}[t]
\includegraphics[width=0.95\linewidth]{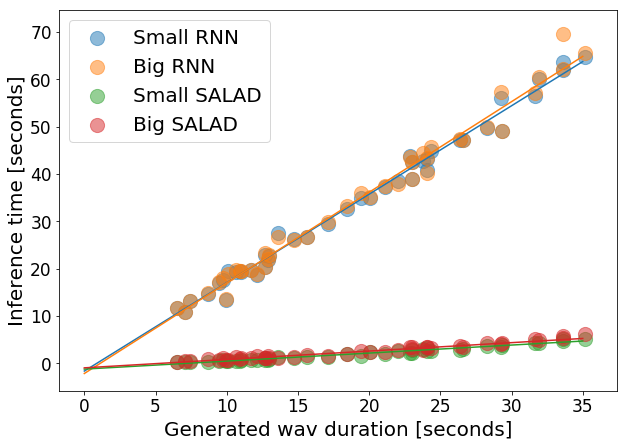}
\caption{\label{fig:cpu_speed} Inference time for the four different models with respect to generated waveform length. Both axis are in seconds.}
\end{figure}

Table~\ref{tab:obj_results} shows the objective results for the systems detailed in section~\ref{sec:model_details} over the two mentioned speakers, M and F. For both speakers, RNN models perform better than the SALAD ones in terms of accuracy and error. Even though the smallest gap, occurring with the SALAD biggest model, is 0.3\,dB in the case of the male speaker and 0.1\,dB in the case of the female speaker, showing the competitive performance of these non-recurrent structures. On the other hand, Figure~\ref{fig:cpu_speed} depicts the inference speed on CPU for the 4 different models synthesizing different utterance lengths. Each dot in the plot indicates a test file synthesis. After we collected the dots, we used the RANSAC~\cite{fischler1981random} algorithm (Scikit-learn implementation) to fit a linear regression robust to outliers. Each model line shows the latency uprise trend with the generated utterance length, and RNN models have a way higher slope than the SALAD models. In fact, SALAD models remain pretty flat even for files of up to 35\,s, having a maximum latency in their linear fit of 5.45\,s for the biggest SALAD, whereas even small RNN is over 60\,s. We have to note that these measurements are taken with PyTorch~\cite{paszke2017automatic} implementations of LSTM and other layers running over a CPU. If we run them on GPU we notice that both systems can work in real time. It is true that SALAD is still faster even in GPU, however the big gap happens on CPUs, which motivates the use of SALAD when we have more limited resources.

\begin{table}[t]
  \caption{Male (top) and female (bottom) objective results. A:~voiced/unvoiced accuracy.}
  \label{tab:obj_results}
  \centering
  \setlength{\tabcolsep}{4pt}
  \begin{tabular}{lcccc}
    \toprule
    \textbf{Model} & \textbf{\#Params} & \textbf{MCD [dB]} & \textbf{F0 [Hz]} & \textbf{A [\%]} \\
    \midrule
    Small RNN                       & 1.17\,M & 5.18 & 13.64 & 94.9             \\
    Small SALAD                     & 1.04\,M & 5.92 & 16.33 & 93.8 		\\
    Big RNN                       & 9.85\,M & 5.15  & 13.58 & 94.9					\\
    Big SALAD					   & 9.66\,M & 5.43 & 14.56 & 94.5       \\
    \midrule
    Small RNN                       & 1.17\,M & 4.63 & 15.11 & 96.8             \\
    Small SALAD                     & 1.04\,M & 5.25 & 20.15 & 96.4 					\\
    Big RNN                       & 9.85\,M & 4.73 & 15.44 & 96.9					\\
    Big SALAD					   & 9.66\,M & 4.84 & 19.36 & 96.6       \\
    \bottomrule
  \end{tabular}
\end{table}

\begin{figure}[t]
\includegraphics[width=0.95\linewidth]{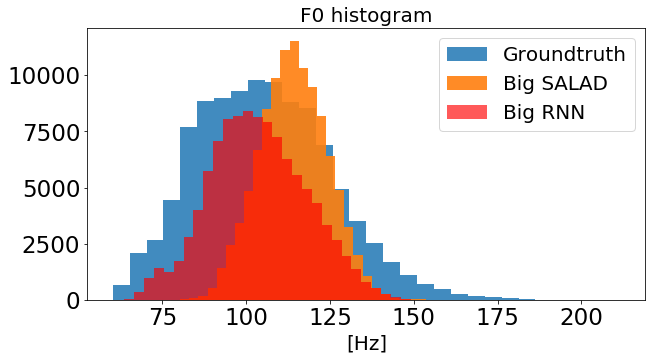}
\caption{\label{fig:f0_hist} F0 contour histograms of ground-truth speech, bigRNN and bigSALAD for male speaker.}
\end{figure}

We can also check the pitch prediction deviation, as it is the most affected metric with the model change. We show the test pitch histograms for ground truth, big RNN and big SALAD in figure~\ref{fig:f0_hist}. There we can see that SALAD's failure is about focusing on the mean and ignoring the variance of the real distribution more than the RNN does. It could be interesting to try some sort of short-memory non-recurrent modules close to the output to alleviate this peaky behavior that makes pitch flatter (and thus less expressive), checking if this is directly related to the removal of the recurrent connection in the output layer. 
Audio samples are available online as qualitative results at http://veu.talp.cat/saladtts . 

\begin{table}[t!]
  \caption{Maximum inference latency with RANSAC fit.}
  \label{tab:ransac}
  \centering
  \setlength{\tabcolsep}{5pt}
  \begin{tabular}{lcc}
    \toprule
    \textbf{Model} & \textbf{Max.~latency [s]} \\
    \midrule
    Small RNN                       & 63.74 \\
    Small SALAD                     & 4.715 \\
    Big RNN                       & 64.84 \\
    Big SALAD					   & 5.455 \\
    \bottomrule
  \end{tabular}
\end{table}

\section{Conclusions}
\label{sec:conclusions}

In this work we present a competitive and fast acoustic model replacement for our MUSA-RNN TTS baseline. The proposal, SALAD, is based on the Transformer network, where self-attention modules build a global reasoning within the sequence of linguistic tokens to come up with the acoustic outcomes. Furthermore, positioning codes ensure the ordered processing in substitution of the ordered injection of features that RNN has intrinsic to its topology. With SALAD, we get on average over an order of magnitude of inference acceleration against the RNN baseline on CPU, so this is a potential fit for applying text-to-speech on embedded devices like mobile handsets. Further work could be devoted on pushing the boundaries of this system to alleviate the observed flatter pitch behavior.

\section{Acknowledgements}
This research was supported by the project TEC2015-69266-P (MINECO/FEDER, UE).


\clearpage
\balance

\bibliographystyle{IEEEtran}

\bibliography{mybib}


\end{document}